\documentclass[conference]{IEEEtran}
\IEEEoverridecommandlockouts
\usepackage{cite}
\usepackage{amsmath,amssymb,amsfonts}
\usepackage{algorithmic}
\usepackage{algorithm}
\usepackage{graphicx}
\usepackage{textcomp}
\usepackage{xcolor}
\usepackage{subfigure}
\usepackage{flushend}
\usepackage{mathtools}
\usepackage{tikz}
\usepackage[normalem]{ulem}
\usetikzlibrary{plotmarks}
\usepackage{algorithm}  
\floatname{algorithm}{Algorithm}
\usepackage{algorithmic}

\def\BibTeX{{\rm B\kern-.05em{\sc i\kern-.025em b}\kern-.08em
    T\kern-.1667em\lower.7ex\hbox{E}\kern-.125emX}}
\begin{document}

\title{Scalable Damper-based Deterministic Networking}

\author{
\IEEEauthorrefmark{1}M. Yassine Naghmouchi,
\IEEEauthorrefmark{2}Shoushou Ren, 
\IEEEauthorrefmark{1}Paolo Medagliani, 
\IEEEauthorrefmark{1}S\'ebastien Martin, 
\IEEEauthorrefmark{1}J\'er\'emie Leguay
\\

\\
\IEEEauthorblockA{Huawei Technologies, \IEEEauthorrefmark{1}\textit{Paris Research Center}, \IEEEauthorrefmark{2}\textit{Beijing Research Center.}}
}

\newcommand{\jeremie}[1]{\textcolor{red}{#1}} 

\maketitle
\thispagestyle{plain}
\pagestyle{plain}

\begin{abstract}
With 5G networking, deterministic guarantees are emerging as a key enabler.
In this context, we present a scalable Damper-based architecture for Large-scale Deterministic IP Networks (D-LDN) that meets required bounds on end-to-end delay and jitter. This work extends the original LDN~\cite{dip} architecture, where flows are shaped at ingress gateways and scheduled for transmission at each link using an asynchronous and cyclic opening of gate-controlled queues. To further relax the need for clock synchronization between devices, we use dampers, that consist in jitter regulators, to control the burstiness flows to provide a constant target delay at each hop. We introduce in details how data plane functionalities are implemented at all nodes (gateways and core) and we derive how the end-to-end delay and jitter are calculated. For the control plane, we propose a column generation algorithm to quickly take admission control decisions and maximize the accepted throughput. For a set of flows, it determines acceptance and selects the best shaping and routing policy. Through a proof-of-concept implementation in simulation, we verify that the architecture meets promised guarantees and that the control plane can operate efficiently at large-scale.

\end{abstract}


\newcommand{\jleg}[1]{\textcolor{red}{#1}}
\newcommand{\sm}[1]{\textcolor{blue}{#1}}
\newcommand{\yn}[1]{\textcolor{violet}{#1}}
\newcommand{\pmed}[1]{\textcolor{orange}{#1}}

\section{Introduction}\label{section1}

Deterministic networking with end-to-end delay guarantees is becoming a must for a wide-range of Internet applications like factory automation, connected vehicles, and smart grids. Indeed, a low and bounded delay (e.g. below 1 $ms$) with nearly zero jitter (e.g. less than tens of $\mu s$) can be vital for some closed-loop control systems related to robots or guided vehicles~\cite{ref_survey}.  
Therefore, networks must not only guarantee deterministic bandwidth for services, but also provide low End-to-End (E2E) delay and jitter, in such a way that data packets are delivered to the destination in time and on time.

In the past decade, a collection of IEEE 802.1 Ethernet standards, known as TSN~\cite{TSNSurvey19}, has been developed to support professional applications over Local Area Networks (LAN) with layer-2 mechanisms such as priority queuing, preemption, traffic shaping, and time-based opening of gates at output ports. To improve the scalability of TSN solutions, the IETF DetNet (Deterministic Networking)~\cite{rfc8655} group has been working on the
Large-scale Deterministic Network (LDN)~\cite{qiang2019large}
architecture
that specifies how traffic should be scheduled and forwarded in large-scale IP networks. 

To address the aforementioned challenges, we presented in~\cite{dip} a comprehensive LDN architecture, that extends current work at IETF~\cite{qiang2019large}, to guarantee deterministic E2E delay and bounded jitter for high-priority flows. In this original LDN  architecture, flows are shaped at ingress gateways (I-GWs) and scheduled for transmission at every hop using a cyclic opening of gate-controlled queues, i.e., IEEE 802.1Qch or Cyclic Queuing Forwarding (CQF), to ensure hop-by-hop guarantees. At the data plane, scalability is achieved thanks to an asynchronous and cyclic opening of $3$ queues and a forwarding of packets based on a hop-by-hop permutation of a header label to identify the next transmission queue. Forwarding operations at intermediate nodes are of low complexity and totally stateless in the core, while flows are shaped at I-GWs to control their arrival rate. At the control plane, an algorithm based on Column Generation (CG)~\cite{desaulniers2006column} has been proposed to decide about acceptance, ingress shaping and routing for each flow. 
While CQF with $3$ (or more) queues enables asynchronous transmissions at each hop and relax the need for an absolute time-synchronization,  synchronization in frequency, i.e., on the duration of each cycle, is still required to ensure that packets are transmitted in the right queue at each hop.

To alleviate the need for synchronization, we consider a LDN architecture based on dampers~\cite{grigorjew2020asynchronous}.  
The concept was introduced by Verma et al.~\cite{verma1991delay}, where a per-flow regulator, called \textit{delay-jitter regulator}, is placed at every node to compensate the time between a target delay, corresponding to the maximum queuing delay of the parent node, and the delay experienced at the previous hop. Cruz~\cite{cruz1998sced+} was the first to use the term damper, which he introduced to guarantee a low deterministic E2E delay and a bounded jitter. 
The main benefit, compared to CQF with $3$ queues, is that it only requires a good accuracy of the oscillator and does not need for frequency synchronization between nodes. Recent works have confirmed that it can work with non-ideal clocks~\cite{mohammadpour2022analysis}. In many practical use cases, such as metro or large enterprise network, nodes do not have time  synchronization capabilities, making an approach like damper applicable and reliable.

This paper extends~\cite{dip} by introducing an efficient damper-based LDN (D-LDN) architecture for both control plane and data plane levels.
The D-LDN network is composed by (i) user devices 
sending and receiving traffic; (ii) edge devices shaping and routing the traffic to enforce isolation between flows and core network devices forwarding flows inside the network,
and (iii) a network controller, responsible for taking admission control decisions. As in LDN, this architecture relies on ingress shaping and stateless forwarding at core nodes. At the data plane level, we present how we implemented dampers using gate-controlled queues, and we detail how end-to-end delay and jitter are calculated. We also analyze the impact that dampers can have on transmission patterns decided at the I-GWs, explaining how this must be taken into account by the controller when taking admission control decisions in order to avoid performance degradation.
To decide which flows to accept and the applied shaping and routing policies, we propose a control plane algorithm that operates efficiently at large scale. It is based on an Integer Linear Programming (ILP) techniques, namely \textit{Column Generation (CG)}~\cite{desaulniers2006column}, used to solve the optimization problem of maximizing the accepted traffic throughput in a D-LDN network~\cite{schrijver2003combinatorial}.

This paper is organized as follows. The state of the art is presented in Sec.~\ref{section:relatedWork}.  In Sec.~\ref{section:dataplane}, we present the data plane mechanisms, including shaping and damper-based forwarding. Sec.~\ref{section:controlplane} is dedicated to the CG-algorithm for the control plane. Sec.~\ref{section:results} provides a proof-of-concept implementation in simulation. Sec.~\ref{section:conclusion} concludes the paper.

\section{Related Work}
\label{section:relatedWork}

Traditionally, data plane scheduling methods rely on round-robin (RR)-based scheduling methods, such as WRR, DRR, MDRR, and URR \cite{WangResearch}, and Weighted Fair Queuing (WFQ)~\cite{GPS, WFQ}. 
In RR-based methods, a system of queues that are scheduled one after another is set up. At each round, an amount of data based on each queue’s weight or deficit is scheduled. This mechanism is used to prioritize traffic, but also to provide deterministic guarantees at each hop with delay bounds. The upper bound of end-to-end delay of these algorithms can be calculated by using the network calculus theory \cite{le2001network}. But, in general, the upper bound of the delay is large and deteriorates greatly with the increase of the number of flows. 
Methods such as Packetized GPS (PGPS) or WF2Q, derived from General Processor Sharing (GPS) \cite{WangResearch}, can guarantee better E2E delay bounds, but they need to maintain per-flow states. Therefore, the traditional scheduling methods have poor scalability and are difficult to be used in large-scale IP networks.

In recent years, 
the IEEE TSN \cite{farkas2018time} working group has proposed a series of standards for Ethernet, including 802.1Qbv, 802.1Qch, 802.1Qcr \cite{specht2016urgency}. 
IEEE 802.1Qch specifies the Cyclic Queuing and Forwarding (CQF) method  \cite{802-1qch} which relies on the use of two cyclic queues. For a given cycle, while one of the queues is being served, the other one is queuing the arriving packets and the cycles change periodically. Although CQF provides low and bounded jitter, it requires time synchronization and can only be used over short distance links. IEEE 802.1 Qcr, also called ATS (Asynchronous Traffic Shaper), specifies methods to manage delay without synchronization between devices. Nonetheless, ATS uses a method that shapes every flow at every hop which lacks of scalability as all core nodes need to maintain per-flow states to be updated for each packet. To integrate TSN technologies into IP networks, the IETF DetNet working group has defined a general DN (Deterministic Networking) architecture~\cite{rfc8655}. 

The main difference between LDN~\cite{dip} (with CQF with 3 or more queues) and D-LDN lies in the mechanism used to schedule packet transmissions. In D-LDN, as each packet carries the information about the time spent in the queues of the parent node, there is no need for strict cycle synchronization between nodes, differently from LDN, in which the cyclic opening of queues must be strictly synchronized. Indeed, each packet must be scheduled for transmission in a specific cycle (i.e., transmission queue) according to the label carried out in the packet header. 
Even if routers may face clock drift, 
mainstream network devices operate 
with a clock accuracy smaller than 100ppm. Thus, 
jitter requirements in the order of  tens of $\mu s$ can be satisfied 
even with non-ideal clock~\cite{mohammadpour2022analysis}.
In addition, our architecture remains functional even in the case of non-ideal clocks \cite{mohammadpour2022analysis}. This makes D-LDN a promising solution for large-scale deterministic IP networks.

Dampers can be quite complex to implement in the data plane. 
Older implementations considered a variant of earliest-deadline-first \cite{verma1991delay, cruz1998sced+} and static priority \cite{zhang1994rate} schedulers. Recent implementations have shown that dampers can coexist with any scheduling mechanism \cite{grigorjew2020asynchronous}. Some of these implementations enforce dampers to behave in a FIFO manner \cite{cruz1998sced+, grigorjew2020asynchronous}. Another similar approach is presented in~\cite{glbf}, where the authors introduce a damper-based forwarding mechanism using an ideal Push-In First-Out Queue (PIFO). Differently from this work, we present how the PIFO can be implemented in the data plane using gate-controlled queues, and we analyze the impact for bandwidth allocation. We also provide an efficient control plane algorithm for admission control.

\section{Data Plane Mechanisms }
\label{section:dataplane}

This section presents the data plane mechanisms at ingress/egress gateways (I-GW/E-GW), and at core nodes.

\subsection{Ingress shaping at I-GWs}
\label{sub:T-r}

According to network calculus~\cite{le2001network}, each flow $f$ can be characterized by an \textit{arrival curve} $A_f(t)= r_f t+b_f$, where $b_f$ is the maximum burst size for $f$, and $r_f$ is the arrival rate. 

Each I-GW shapes incoming flows with bursts $b_f$ into smaller bursts of size $b'_f$. 
I-GWs implement shaping via the asynchronous cyclic opening of Gate Controlled Queues (GCQs), in order to schedule packet transmissions, and whose utilization allows supporting  \textit{transmission patterns}, i.e. the mapping of incoming flows into specific GCQs reservations, over a hypercycle of length $HC$ cycles, each of them of duration $T$. As only one cycle is active at the same time, the corresponding queue opens for transmission at the beginning of the  cycle and closes at the end. The remaining of the time, the queue is open for reception and closed for transmission. 

\begin{figure}
    \centering
    \includegraphics[width=1\columnwidth]{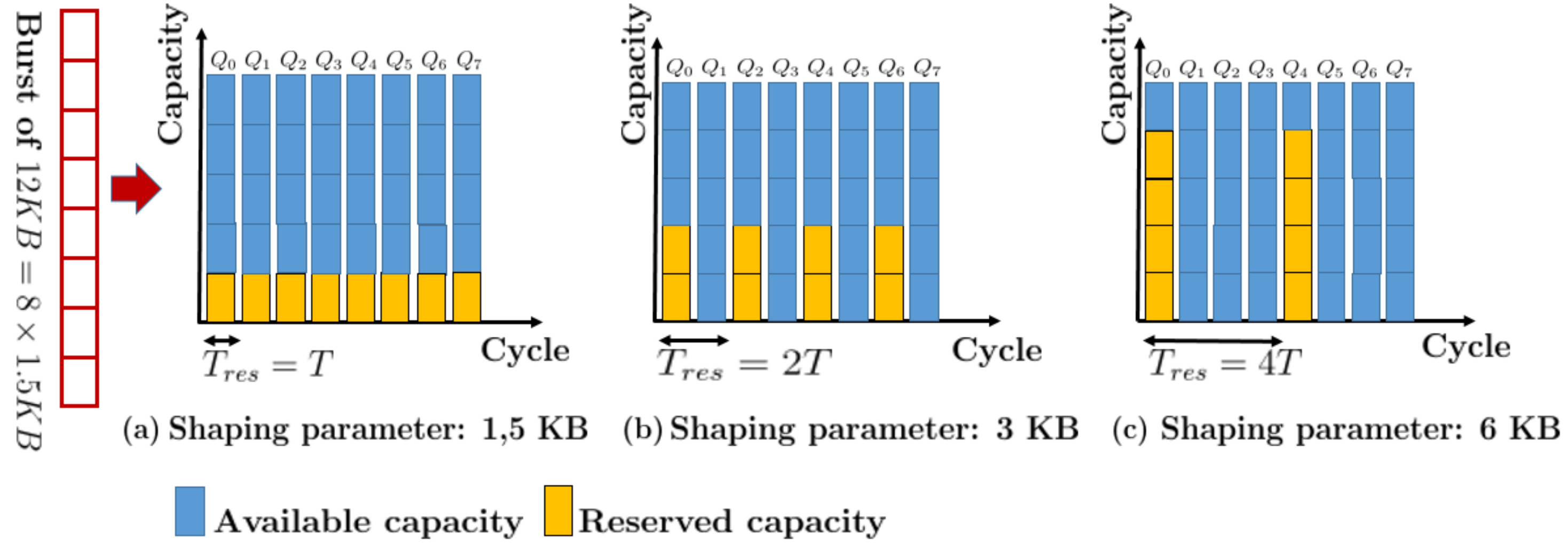}
    \caption{Examples of reservation patterns to shape an incoming flow at I-GW.}
    \label{fig:shaping}
\end{figure}

Incoming flows are injected at I-GWs into transmission patterns selected by the controller. In this paper, we consider \textit{regular} patterns, where reservations in transmission queues are of the same amount of resources and separated with a constant period $T_{res}$. When $T_{res}=T$, the reservation is \emph{uniform} and equal at all cycles.
As an incoming burst is spread over $\lceil\frac{b_f}{b'_f}\rceil$ cycles, every $ T_{\rm res}$ cycles, the shaping introduces an additional delay of $d_{shaping} = T_{\rm res} \lceil\frac{b_f}{b'_f}\rceil$, which is the \textit{shaping delay} experienced by the last packet of a maximum burst size. 
Fig.~\ref{fig:shaping} shows three possible reservations for regular patterns at the I-GW, for an incoming flow with a burst of $12$ KB and a maximum packet size of $1.5$KB.

\subsection{Damper-based Forwarding}
\label{Sec:fowarding}

While several damper architectures~\cite{shou44damper} are possible, we will consider in this paper the architecture of Fig.~\ref{fig:forwardingProcess} which illustrates a damper pair that consists of three elements: 1) the queuing system of the parent node $h$, 2) the transmission by the parent node $h$, and the reception by the child node $h+1$, and 3) the damper module on the child node $h+1$, used to compensate the queuing delay experienced at  parent node $h$. This approach allows keeping the node architecture fairly simple, as a single damper is used. 

\subsubsection{Scheduling with dampers}

\begin{figure}[t]
    \centering
    \includegraphics[width=1\columnwidth]{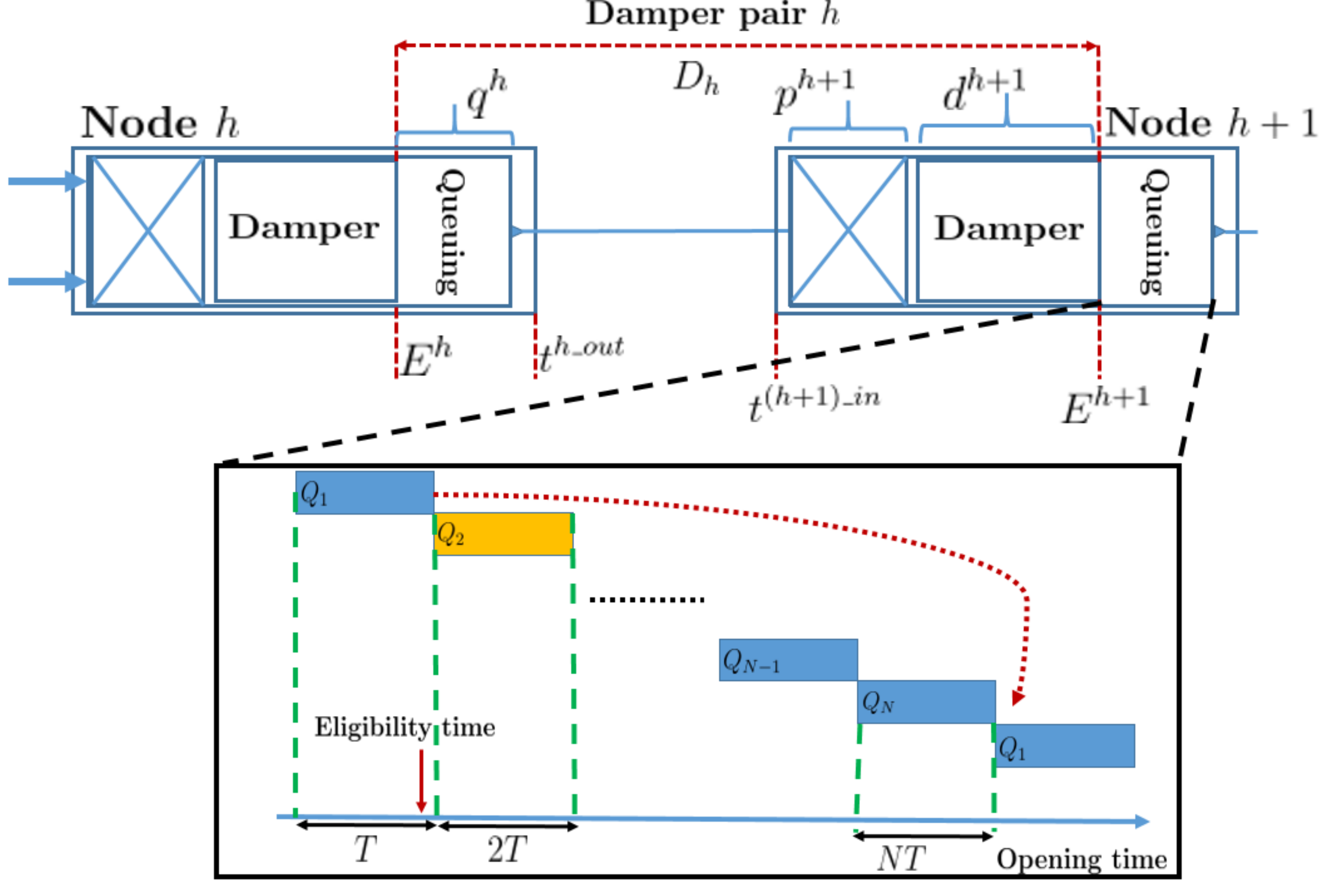}
    \caption{Damper-based forwarding between nodes $h$ and $h+1$.}
    \label{fig:forwardingProcess}
\end{figure}

In our architecture, packet scheduling is implemented via a queuing system at each port, represented at the bottom of Fig.~\ref{fig:forwardingProcess} and operating as follows:  for a system of $N$ queues, the queue $Q_{i-1}$ opens before the queue $Q_i$. We point out that only one queue can be open at the same time. The time interval between the opening of $Q_{i-1}$ and $Q_{i}$ is fixed and equals to $T$.  
When a packet arrives at the node $h$, the damper mechanism is responsible for deciding in which queue a packet must be transmitted, in order to let it experience the same delay $Q$ as the other packets of the same flow. For this reason, given the information carried in the header of the packet about the time $q^h$ spent in the parent node $h$, the child node $h+1$ can compute at which time the packet must be released by the damper and select the queue accordingly. This time instant $E^{h+1}$ is referred to as \textit{eligibility time}. According to our implementation, a packet will be inserted in the next queue opening after the eligibility time, resulting in a maximum queuing delay $Q=2T$. This bound comes from the worst case analysis, according to which a packet is queued in $Q_i$ just after the opening of $Q_{i-1}$ and has to wait until the end of $Q_{i-1}$ for being transmitted.

\subsubsection{Forwarding Process}
The damper module must enforce a constant delay $D^h$ for all packets passing through the damper pair.  The expression for $D^h$ is given by Eq.~\ref{DamperDelay}, where $q^h$ is the actual queuing delay, $p^{h+1}$ is the actual processing delay and $d^{h+1}$ is the time a packet needs to wait in the damper. $Q^h$ and $P^{h+1}$ represent respectively upper bounds on the queuing and processing delays.  
\begin{equation}
\label{DamperDelay}
    D^h=q^h+p^{h+1}+d^{h+1} = Q^h+P^{h+1}
\end{equation}
In Eq.~\ref{DamperDelay}, we suppose that the processing delay is constant: $p^{h+1} = P^{h+1}$. Besides that, since the propagation  delay is nearly constant in wired networks, it is not considered in Eq.~\ref{DamperDelay}.

The packet forwarding process is described as follows, assuming that a packet has already been put into the damper by node $h$, that the next hop is determined using standard IP routing (and determined by the network controller), and the eligibility time $E^h$ has already been computed for the node $h$:
\begin{enumerate}
    \item Node $h$ enqueues the packet for transmission in the first available queue following the eligibility time $E^h$;
    \item When the queue is opened for transmission, packets inside are sent in a FIFO order to the next hop, selected either via Segment Routing (SR) labels or via routing tables;
    \item Node $h$ computes the time spent in the queues by the packet as $q^h=t^{h\_out}-E^h$ and encapsulates it in the header of the packet, together with the information about the expected duration $Q^h$;
    \item Node $h+1$ receives the packet and process it, experiencing a processing delay $P^{h+1}$. 
    \item The damper module of $h+1$ computes the eligibility time $E^{h+1}$.
\end{enumerate}
The steps described above are repeated for each core node that forwards packets, including the E-GW that sends data to the final client. We point out that at the I-GW, as there is no parent node sending packets,  $E_0$ is set to the first opening queue matching with the selected transmission pattern.

The eligibility time $E^h$ at the hop $h+1$ can be expressed as follows:
\begin{equation}
    \label{EligibilityTime}
    E^{h+1}=t^{(h+1)\_in} + P^{h+1} + [Q^h - t^{(h)\_out} +  E^h],
\end{equation}
where $t^{(h+1)\_in}$ is the time at which node $h+1$ receives the packet, and $t^{(h)\_out}$ is the time at which node $h$ sends the packet.

\subsubsection{Damper Impact on Traffic}
\label{damperimpactsection}
The damper, in this implementation common to all the ports of a device, is introduced to compensate the queuing delay experienced by a packet in the parent node, in order to provide a constant forwarding delay for all the packets of the same flow.

\begin{figure}[!t]
    \centering    \includegraphics[width=1\columnwidth]{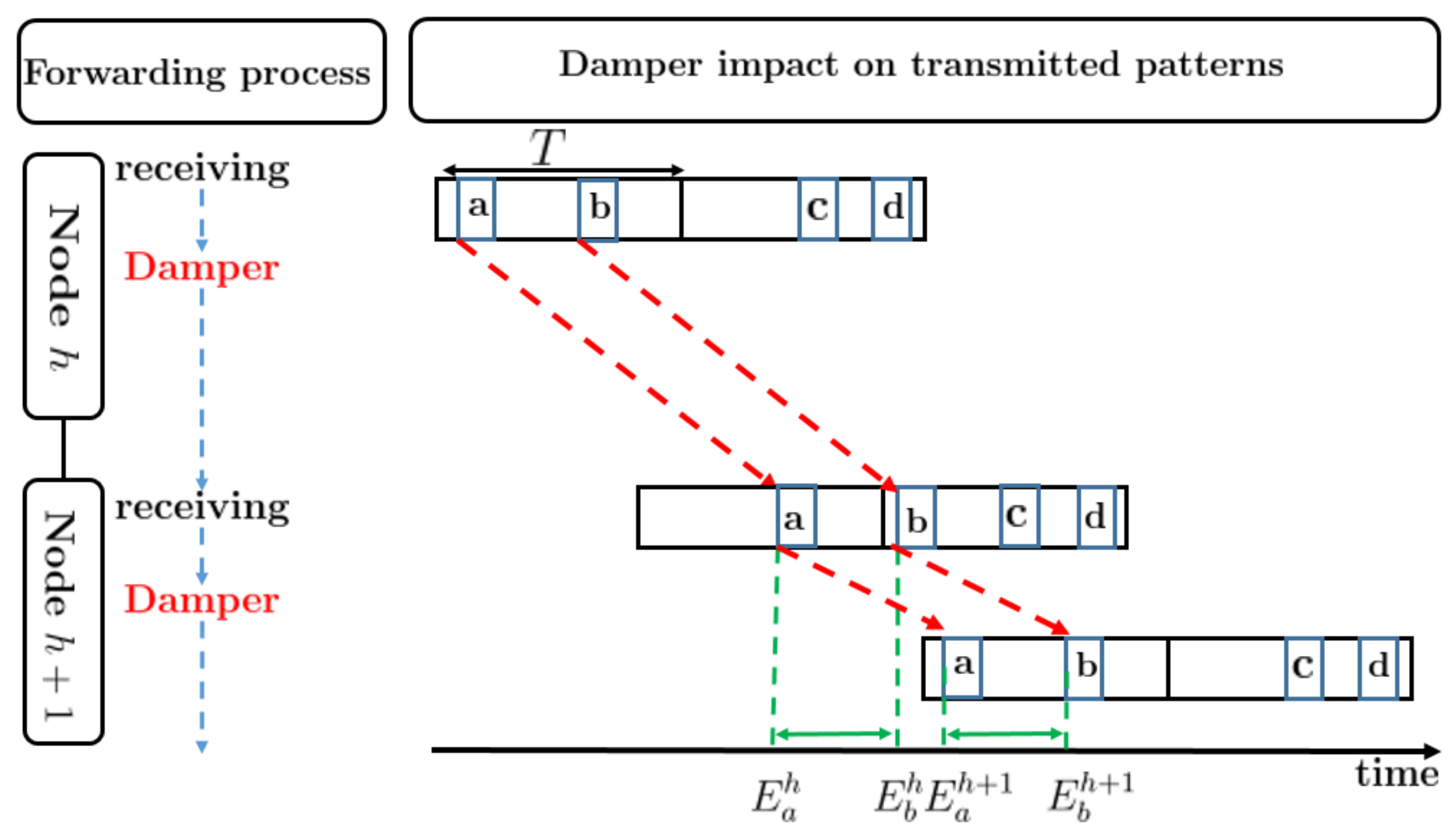}    \caption{Impact of damper on a transmitted pattern.}
    \label{fig:damperimpact}
\end{figure}

 Neglecting the fixed term $P^{h+1}$, as $Q^h=q^h_b + d^{h+1}_b=2T$, at its eligibility time $E^h_b$ in the node $h$, a packet $b$ can be enqueued either in the first opening queue (if $q^h_b \leq T$), or in the second opening queue (if $q^h_b \geq T$). Therefore, it may undergo a shift delay of one additional cycle.
 The delay experienced by a packet in the damper can then change the transmitted pattern structure initially selected by the I-GW through shaping. An example is given in Fig.~\ref{fig:damperimpact}. Here, the packet $b$, received by the node $h$ in cycle 1, is delayed by the damper to be transmitted in cycle 2 together with packets $c$ and $d$. 
 
From Eq.~\ref{EligibilityTime}, we can deduce that:
\begin{equation}
    \label{EligibilityTimeGap}
    E^{h}_a - E^{h}_b =  E^{h+1}_a - E^{h+1}_b \leq T,
\end{equation}
where $h$ and $h+1$ is a damper pair, and $a, b$ are two packets of the same flow transmitted in the same cycle, as shown in Fig.~\ref{fig:damperimpact}.
Eq.~\ref{EligibilityTimeGap} says that the eligibility time gap between any pair of packets transmitted in the same cycle remains constant in any node of the path, and it is not larger than $T$. This guarantees that the delay introduced by the dampers over any E2E path is at most one additional cycle. 
In the example of Fig.~\ref{fig:damperimpact}, we can see that  the packet $b$ is received by the node $h+1$ in the second cycle. The worst case scenario occurs when the damper $h$ delay leads to a transmission of the packet $a$ in the second cycle. 

From a control plane point of view, when deciding about bandwidth allocation, we must take into account that due to the damper, in the worst case all the packets of two adjacent cycles can be transmitted on the same cycle and the next hop must have the sufficient capacity to accept both of them. This configuration occurs when the pattern period $T_{res}$ is equal to the cycle period $T$. However, if $T_{res}>T$, there is no risk that packets from adjacent cycles will overlap on the same cycle. 
\textit{The maximum pattern reservation} on a cycle for a flow $f$ and an initial reservation pattern $k$ with a shaping parameter $b'_f$ is denoted by $\beta(f,k)$ and given by

\begin{equation}
    \label{maxres}
    \beta(f,k) = \left\{
    \begin{array}{ll}
        2b'_f & \mbox{if } T_{res}=T,\\
         b'_f & \mbox{otherwise.}
    \end{array}
\right.
\end{equation}

\subsubsection{E2E delay and jitter bounds}

Over a path composed of $H$ devices, the upper delay bound (excluding link propagation delays) can be written as

\begin{equation}
\label{endtoenddelay}
    D^{E2E}= \sum\limits_{k=1}^H D^k.
\end{equation}

Damper guarantees that the delays $D^1, \ldots, D^{H-1}$ experienced by any packet at each pair are the same. However, the damper at the node $H$ constitutes an incomplete pair, and we have $D^H=Q^H$, which only accounts for the maximum queuing delay.  
Due to the damper mechanism, nodes from $1$ to $H-1$ do not generate jitter and the E2E jitter along the path comes only from node $H$. Hence, the E2E jitter bound is $Q^H=2T$.

\section{Control Plane Algorithm}
\label{section:controlplane}
Here, we formally define the admission control problem that the control plane needs to solve. We present a path-based formulation and a CG-based algorithm to maximize the accepted traffic throughput over the D-LDN network. 

\subsection{Problem Statement}
An instance of the  problem is given by a couple $(G, \mathcal{F})$: 
\begin{itemize}
    \item $G=(V,A)$ is a digraph representing the network topology, where $V$ is the set of network devices and $A$ is the set of arcs representing  physical links. Each node $v \in V$ has a buffer capacity $c_v$ shared over all ports (in data units). Each arc $a=(i,j) \in A$ has a transmission delay $l_a$  and a link capacity $c_a$ (in data units). We have $l_a=D^i+Prop_{ij}= Q^i+P^j+Prop_{ij}$, where $Prop_{ij}$ is the propagation delay from $i$ to $j$.

    \item $\mathcal{F}$ is a set of flows that need to be admitted with proper shaping and routing policies. Each flow $f \in \mathcal{F}$ is characterized by:
    \begin{itemize}
        \item a source $s^f \in V$ and a destination $t^f \in V$;
        \item an arrival curve $A_f(t)=r_ft+b_f$, where $r_f$ is the flow rate and $b_f$ is the maximum burst size;
        \item a throughput $R_f$;
        \item a maximum end-to-end delay $D_f$; 
        \item a set of possible transmission patterns $\Pi^f$ such that each flow $f$ of pattern $k$ has a period $T(f,k)$, a maximum per cycle reservation $\beta(f,k)$, defined by Eq.~\ref{maxres}, and a shaping delay $d(f,k)$.
    \end{itemize} 
\end{itemize}

For every flow $f \in \mathcal{F}$, let $\mathcal{P}^f$ denote the set of paths between $s^f$ and $t^f$. Let us also denote $S_f$ the set of path-pattern couples such that the end-to-end delay constraint is respected. It is formally defined as follows:   $$\mathcal{S}_f=\{(p, k): p \in \mathcal{P}^f, k \in \Pi^f \mbox{ and }\sum\limits_{a \in p} l_a + d(f,k) \leq D_f \}.$$

A feasible solution to the problem consists in selecting for each flow at most one element in $\mathcal{S}_f$, i.e., select or not for each flow a single pattern in $\Pi^f$ and a single path in $\mathcal{P}^f$, respecting the end-to-end delay constraints, in such a way that the following constraints are satisfied: 
\begin{itemize}
    \item[(1)] \textit{Arc-capacity constraints}: for each arc $a \in A$, the sum of $\beta(f,k)$, the maximum per cycle reservations for each flow $f$ of pattern $k$ passing through link $a$, does not exceed  capacity $c_a$;
    \item[(2)] \textit{Buffer capacity constraints}:  for each node $v \in V$, the sum of $\beta(f,k)$, the maximum per cycle reservations, for each flow $f$ of pattern $k$ crossing node $v$, does not exceed the buffer capacity  $c_v$;
\end{itemize}

The overall objective of the admission control problem is to maximize the total accepted throughput. 
The capacity constraints (1) and (2) are defined under a worst-case scenario for the way transmission patterns of multiple flows can combine at each node or link.
It has the advantage of being simple and robust against the worst case.

\subsection{Mathematical Model}
The problem is equivalent to the following path-based ILP Damper Formulation ($DF$): 
\begin{alignat}{4}
\nonumber \textbf{(DF)} & \max \sum_{f \in \mathcal{F}}\sum\limits_{s \in \mathcal{S}_f} R_f x_{f,s} & \quad & \tag{constraint: dual variables} \\
\label{path:routing} & \sum\limits_{s \in \mathcal{S}_f} x_{f,s} \leq 1 & \quad & f \in \mathcal{F}, \tag{1. routing and shaping: $\lambda_f$} \\
\label{path:arc_capacity} &   \sum_{f \in \mathcal{F}}\sum\limits_{s=(p,k) \in \mathcal{S}_f: a \in p}  \beta(f,k) x_{f,s} \leq  c_{a}& \quad &  a \in A, \tag{2. arc capacity: $\mu_{a}$} \\
\label{path:txb_capacity} & \sum_{f \in \mathcal{F}}\sum\limits_{s=(p,k) \in \mathcal{S}_f: v \in p}   \beta(f,k) x_{f,s} \leq  c_{v} & \quad & v \in V,  \tag{3. buffer capacity: $\omega_{v}$}\\
\label{path:integrity} &  x_{f,s} \in \{0,1\}   & \quad &   f \in \mathcal{F}, s \in \mathcal{S}_f  \tag{4. integrality}
\end{alignat}

Constraints (1) are routing constraints, they ensure that  each accepted flow has exactly one path-pattern couple. Constraints (2), and (3) are respectively link capacity and buffer capacity constraints. Finally, (4) are integrality constraints. 
The number of constraints in (DF) is polynomial: $|F|+|A|+|V|$. As the number of paths in general graphs is exponential, the number of variables in (DF), namely \textit{columns}, may be exponential. Therefore, it is not possible in general to solve the entire problem with a solver. However, we can use a column generation algorithm with an ILP Rounding procedure to obtain high quality  solutions. The algorithm we develop is called \textit{CGX}, and it is described in the following. 

\subsection{CGX Algorithm}
 The Column Generation with exact (CGX) rounding first solves the \textit{linear relaxation} $LDF$ of (DF), which relaxes the integrality constraints (4) on the variables $x_{f,s}$, and then rounds the LDF solution to an integer solution, using an ILP solver. The overall algorithm is described in Algorithm~\ref{Offlinealgo}. The optimal solution to the linear relaxation provides an Upper Bound (UB) to DF, and it can be used to evaluate the optimality gap. Knowing an integer solution of objective value $Z$, the optimality gap is given by $\frac{UB-Z}{UB} \times 100$.

\subsubsection{Solving the linear relaxation}
It is well-known that Linear Programs (LPs) such as LDF can be solved in polynomial time in terms of input size \cite{khachiyan1979polynomial}. However, the number of variables of $LDF$ is not polynomial. We overcome this problem by applying column generation, which permits to obtain an optimal solution to the linear relaxation $LDF$ by generating only a polynomial subset of variables. 

The column generation procedure starts with a \textit{restricted master} Linear Program (LP) $LDF^0$; with no variables for our case. By solving \textit{the pricing problem}, a method based on LP duality \cite{schrijver2003combinatorial}, we decide whether there are variables that are currently not contained in the restricted master LP, but that might improve the objective value. If no such variables can be found, the current subset of variables is guaranteed to be sufficient to solve the \textit{master} $LDF$ problem optimally. Otherwise, newly generated variables are added to the restricted LP and the process iterates. 

For a given iteration $i$, let $LDF^i$ and $\mathcal{S}^i_f$  denotes respectively the restricted linear relaxation and its associated set of columns indices. 
Remark that a solution $x^i_{f,s}$ of $LDF^i$ induces a feasible solution $x_{f,s}$ of LDF by setting $x_{f,s}=x^i_{f,s}$ for all $f \in \mathcal{F}, s \in \mathcal{S}^i_f$ and $x_{f,s}=0$ otherwise. 
We can determine that the induced solution $x_{f,s}$ is optimal for LDF by considering  $D-LDF^i$, the dual of the $LDF^i$:

\begin{alignat}{4}
\nonumber  & \min \sum\limits_{f \in \mathcal{F}} \lambda_f +\sum\limits_{a \in A} c_{a} \mu_{a}+ \sum\limits_{v \in V} c_v \omega_{v}\\
\nonumber & \sum\limits_{a=(u,v) \in p}  \beta(f,k) ( \mu_{a} +\omega_{v} ) \geq  R_f -  (\lambda_f+\beta(f,k) \omega_{s^f}), \\
\nonumber &   f \in \mathcal{F}, s=(p,k) \in \mathcal{S}^i_f,  ~~~~~~~~~~~~~~~~~~ \mbox{(Dual constraint)} \\
\nonumber &  \lambda_f, \mu_{a},  \omega_{v}  \geq 0  ~~~~~~~~~~~~~~~~~~~~~~~~  f \in \mathcal{F}, v \in V, a \in A. 
\end{alignat}

Let $(\lambda^*_f, \mu^*_{a}, \omega^*_{v})$ be the optimal solution of $D-LDF^i$. if there exists a flow $f$, a pattern $k$ and a path $p$ such that 
\begin{align}
 \label{dualpath:pp} \sum\limits_{a=(u,v) \in p}  \beta(f,k) ( \mu^*_{a} +\omega^*_{v} ) <  R_f -  (\lambda^*_f+\beta(f,k) \omega^*_{s^f}).
\end{align}
then, the solution is infeasible to $D-LDF^i$. The problem $D-LDF^{i+1}$ with $\mathcal{S}^{i+1}_f = \mathcal{S}^i_f \cup s^{i+1}$ is an improved approximation to $D-LDF^i$, where $s^{i+1}=(k,p)$. If no such path exists, the solution is feasible to $D-LDF$ and also optimal to $LDF$.

The pricing problem consists in finding a separating path for each flow-pattern couple. This reduces to solving a Constrained Shortest Path (CSP) \cite{aneja1978constrained} problem in the graph $G$. The goal is to find a path $p^*$ that minimizes $\sum\limits_{a=(i,j) \in p^*}\beta(f,k) ( \mu^*_{a} + \omega^*_{j})$ while respecting the delay constraints. 

\subsubsection{ILP Rounding}
In this final step, we consider the linear relaxation $LDF^*$ of the last column generation iteration and reapply integrality constraints (4). Using an ILP solver, we obtain a feasible solution to the original (DF) problem. 

\begin{algorithm} [t]
\begin{small}
\caption{CGX: Column Generation and ILP Rounding}
\begin{algorithmic} \label{alg:CGRR}
\REQUIRE An instance $(G, \mathcal{F})$ of the problem. 
\WHILE{Columns added or first iteration}
\STATE Solve the restricted LP and get the dual variables values
\FORALL{flow $f \in \mathcal{F}$, pattern $k \in \Pi^f$}
\STATE Path $p^* \leftarrow \emptyset$. 
\STATE Solve pricing problem for $f$ and $k$ 
\IF {path $p^*$ found such that inequality (\ref{dualpath:pp}) is satisfied} 
\STATE add the variable $x_{f,(k,p^*)}$ to the restricted LP
\ENDIF 
\ENDFOR
\ENDWHILE
\STATE Set the columns added to 0-1 
\STATE Solve the ILP using an ILP solver
\RETURN Solution of the ILP
\end{algorithmic}
\label{Offlinealgo}
\end{small}
\end{algorithm}

\section{Performance Evaluation}
\label{section:results}

We now verify with a proof-of-concept implementation in simulation that the D-LDN architecture can guarantee E2E delay requirements and jitter bounds at large-scale. We also show the efficiency of CGX algorithm in terms of accepted throughput and computational time. 

\subsection{Proof-of-concept Implementation}

We conduct simulations results based on a scenario of 5 flows where the egress bandwidth of all devices is 10 Gbit/s. The average rate of flow 1 is 2.24 Gbps, and the maximum burst size is 1400 bytes. The average rate of flows 2 and 3 is 6.72 Gbit/s, and the maximum burst size is 4200 bytes. The average rate of flow 4 and flow 5 is 3.36 Gbps, and the maximum burst size is 2100 bytes. In addition, some best-effort traffic passes through each hop, and low-priority traffic is used. The upper bound $Q$ of the queuing delay of each hop interface is set according to the network calculus theory, and it is $5\mu{s}$, and the transmission delay is omitted.

\begin{figure}[h]
    \centering
    \includegraphics[width=1\columnwidth]{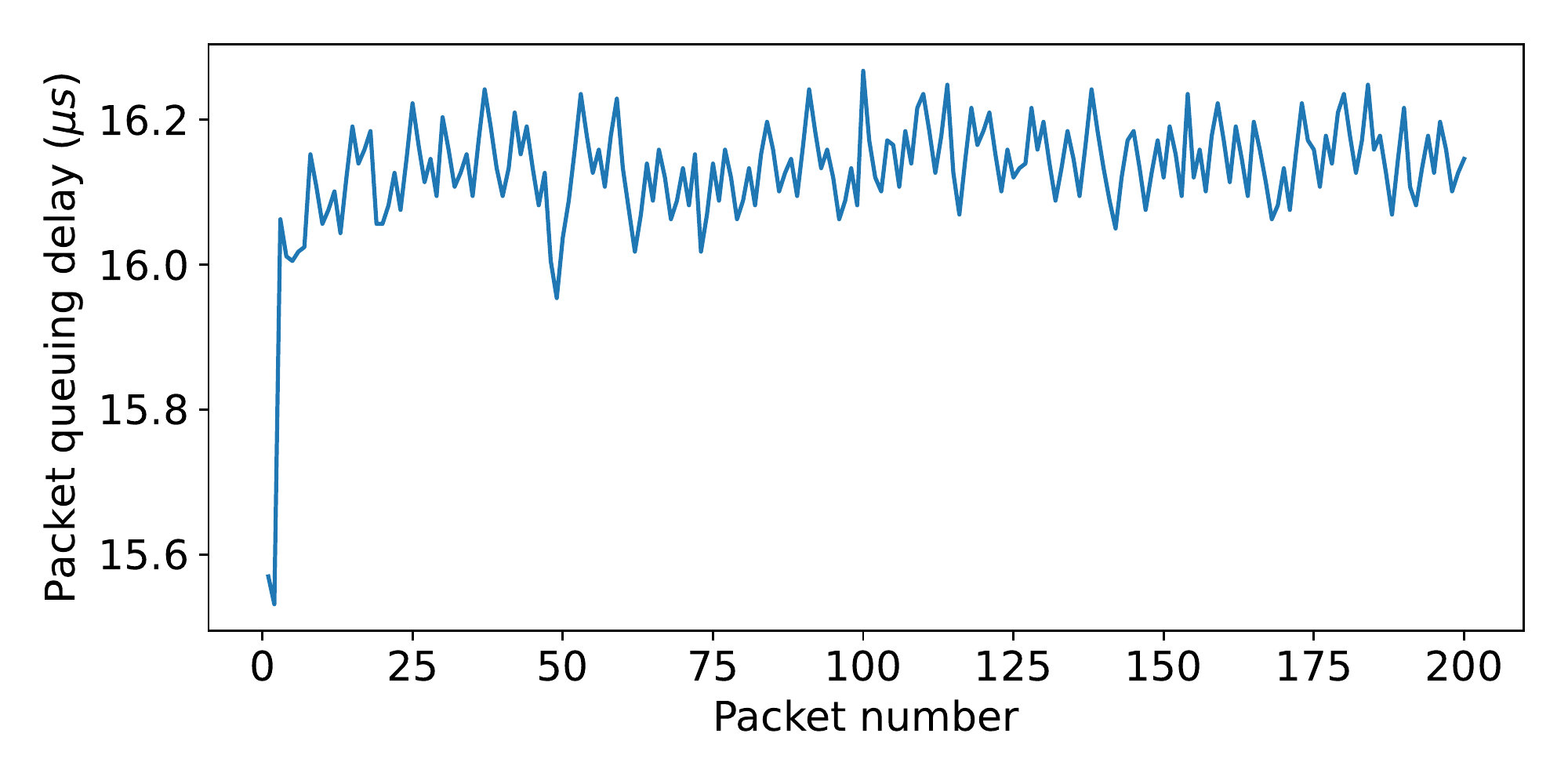}
    \caption{E2E Queuing Delay }
    \label{fig:queuingdelay}
\end{figure}

We collect statistics on the E2E queuing delay of packets for all flows. Fig.~\ref{fig:queuingdelay} shows the results for flow 1. As we can see, even when an interfering flows exist, based on the damper scheduling mechanism, the D-LDN network can still be ensured that the E2E delay is respected, and the jitter does not exceed the worst delay of the last hop $Q=5\mu{s}$.

\subsection{Admission Control}
\subsubsection{Instances description and implementation features}

We have generated a realistic network topology composed of $505$ nodes and $1061$ links, each with a maximum propagation delay of $40\mu{s}$.
Based on this topology, we build two families of instances. The first one is obtained by varying the capacity of links and  nodes at \emph{capacity levels} from 1 to 10 and with an E2E delay requirement of $1{ms}$ for all flows. At capacity level $i$, instances correspond to a maximum buffer capacity of  $i \times 10 Mb$ and maximum link capacity of $i \times 100 Gb/s$, respectively.  
The second family of instances is obtained by considering the capacity level $10$ and varying the E2E delay requirements of flows in $100, 200,\ldots, 1000\mu{s}$. 
In all families of instances, for a given capacity level and a given E2E delay requirement, we also vary the number of flows in $100, 500, 1000, \ldots, 5000$ and select origin/destination pairs at random. In a simulation, all flows have the same  E2E delay requirement, the same maximum burst size of $1500$ bytes, a random throughput in $1, \ldots, 10$ $Gb/s$. At I-GWs, we consider a hypercycle length $HC$ of $8$ cycles, and we set the cycle duration to $T=10\mu{s}$.

We implement our algorithms in C++ and solved the linear programming formulation using the  CPLEX 12.6 \cite{cplx} solver. As we are dealing with an off-line planning algorithm, we limit the overall resolution time of CGX to $5$ mins.

\subsubsection{Numerical results}
We compare the performance of CGX  to the OSPF routing protocol, which selects the shortest path for each flow where each link $a \in A$ has a cost of $\frac{10^8}{8 \times c_a}$. To this end, we introduce the accepted throughput gap as $ \frac{Th(CGX)}{Th(OSPF)} \times 100,$ where $Th$ is the accepted throughput. 

\begin{figure*} [!ht]
	\centering
	\subfigure[Accepted throughput gap between CGX and OSPF.]{
		\includegraphics[width=0.45\textwidth]{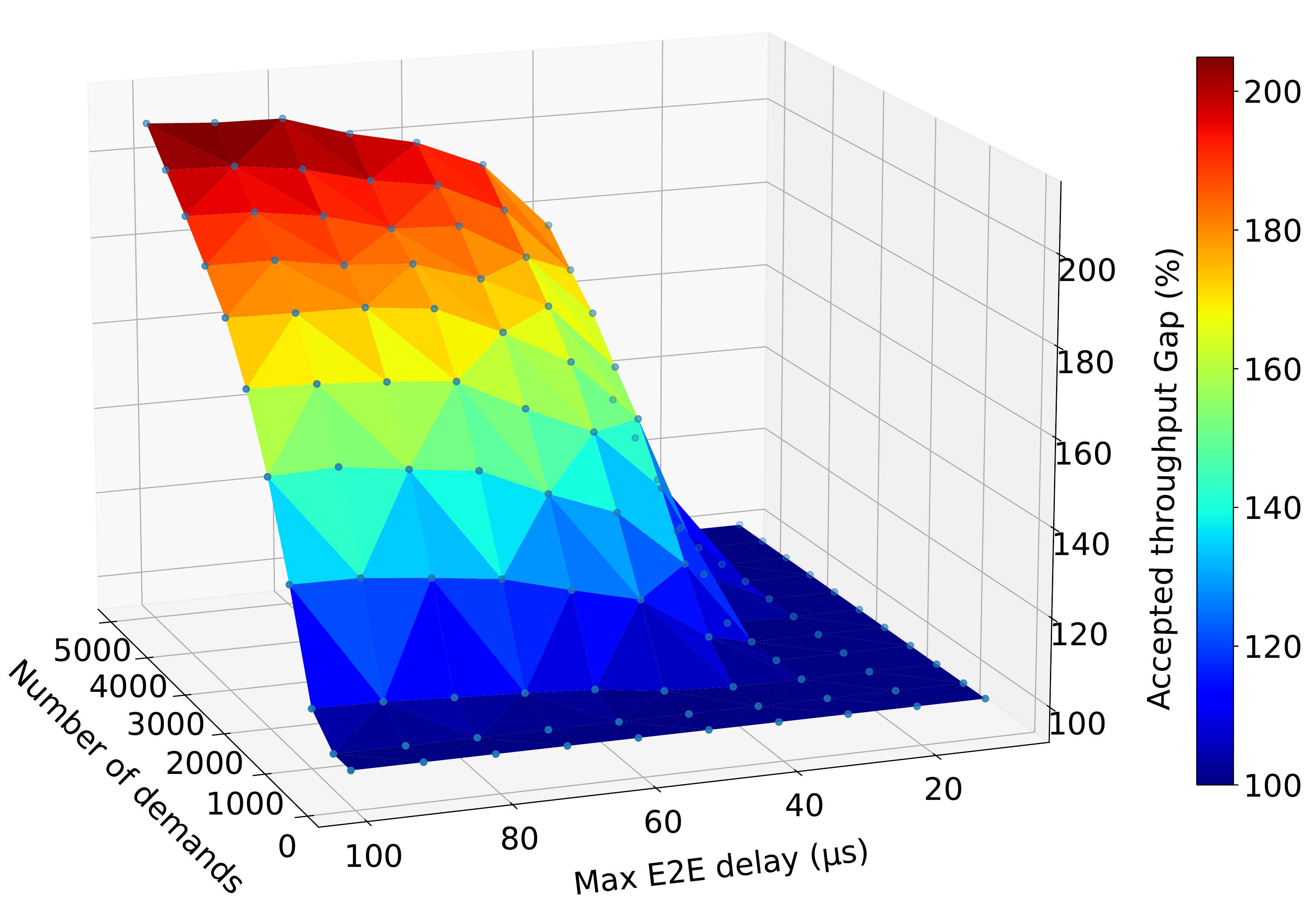}
		\label{fig:gapdelay}
	}
	\subfigure[Computational time for CGX (s).]{
		\includegraphics[width=0.45\textwidth]{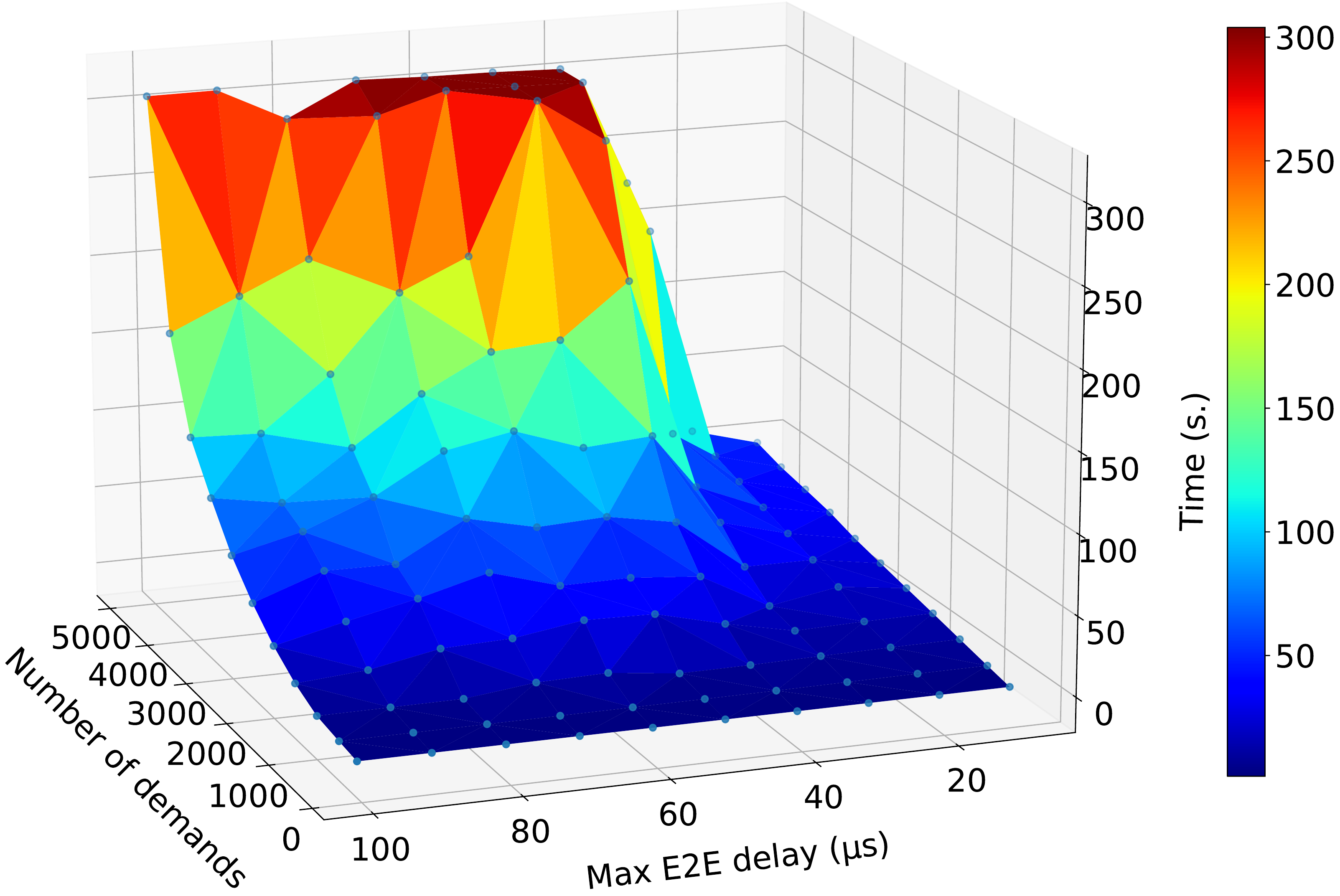}
		\label{fig:timedelay}
	}
	\caption{Admission control results for CGX and OSPF routing: sensitivity to E2E delay requirements and number of demands.}

	\label{fig:sensitivity}
\end{figure*}

Fig.~\ref{fig:gapdelay} reveals the sensitivity of  the accepted throughput gap between CGX and OSPF to the maximum E2E delay and the number of demands. As we can see, for some demands or for a weak E2E maximum delay, the throughput accepted by CGX and the one accepted by OSPF are the same. In fact, when the number of demands is smaller than $1000$ or the E2E delay is smaller than $20 \mu s$, the gap is equal to $100\%$.
Nonetheless, we can notice that this gap increases with the increase of the number of demands and the maximum E2E delay to reach $200 \%$ for a maximum E2E delay between $80 \mu s$ and $100 \mu s$ and a number of demands greater than 4500. In that case, corresponding to large instances, CGX accepts 2 times more traffic throughput than OSPF. To conclude, CGX solutions are clearly of better quality than OSPF.

Fig.~\ref{fig:timedelay} shows the evolution of the solution time for CGX, according to E2E delay. We can see that the time limit of $5$ mins is reached only with $4500$ and $5000$ demands. For this number of demands, the resolution time is sensitive to the E2E delay requirements. Indeed, we can notice in Fig.~\ref{fig:timedelay} a decrease of the resolution time for low or very high delays (10, 90, 100 $\mu s$,). Finally, note that the solutions of our algorithm also have the merit of being optimal or almost optimal. In fact, for all the instances that we have tested with a time limit of $5$ min, CGX has an average optimality gap of $0.35$\%.

\section{Conclusion and Perspectives}
\label{section:conclusion}

We have proposed D-LDN, a damper-based architecture for large-scale deterministic networks with End-to-End delay and bounded jitter guarantees. We have introduced the data plane mechanisms and the theoretical background to determine E2E delay and jitter bounds. To maximize traffic throughput acceptance in the network, we have developed an efficient control plane algorithm called CGX based on column generation. Through a proof-of-concept implementation we verified that D-LDN can meet strict QoS guarantees. We also presented numerical results to demonstrate that our CGX algorithm gives very good solutions on large-scale instances.

The control plane algorithm we have introduced in this paper is an offline algorithm. An interesting direction of this work is to study and design an efficient online algorithm to solve the admission control problem. 

\bibliographystyle{IEEEtran}
\bibliography{bibliography}

\end{document}